\documentclass[aps,pra,twocolumn,showpacs]{revtex4}
\usepackage{eurosym}
\usepackage{amsfonts}
\usepackage{amsmath}
\usepackage{amsopn}
\usepackage{amssymb}
\usepackage{graphicx}
\usepackage{graphics}
\usepackage{color}
\usepackage{verbatim}

\begin{document}

\title{Regeneration of Airy pulses in fiber-optic links with dispersion management of the two leading dispersion terms of opposite signs}
\author{R. Driben$^{1}$ and T. Meier$^{1}$}
\affiliation{$^{1} $Department of Physics and CeOPP, University of Paderborn, Warburger Str. 100, D-33098 Paderborn, Germany}
\date{\today}

\begin{abstract}
Dispersion management of periodically alternating fiber sections with opposite signs of two leading dispersion terms is applied for the regeneration of self-accelerating truncated Airy pulses. It is demonstrated that for such a dispersion management scheme, the direction of the acceleration of the pulse is reversed twice within each period. In this scheme the system features light hot spots in the center of each fiber section, where the energy of the light pulse is tightly focused in a short temporal slot. Comprehensive numerical studies demonstrate a long lasting propagation also under the influence of a strong fiber Kerr nonlinearity.
\end{abstract}

\maketitle

\section{Introduction}

In recent years the interest in fascinating self-accelerating Airy light waves has increased significantly due to their unusual fundamental properties and their application potential \cite{1,2,3,4,5,6}. While other types of waves such as nonlinear solitary waves  may experience an acceleration (central frequency shift) as a result of interaction with other waves \cite{7}, the acceleration of Airy light waves is its inherent property. The Airy beams have a multi-peak shape and propagate along bending trajectories. Airy pulses \cite{8,9,10,11,12,13,14} - counterparts of spatial Airy beams - propagate with their acceleration resulting from a varying group velocity. Studies of multi-dimensional spatio-temporal Airy light bullets were also presented  \cite{12,15}. Applications of Airy waves include plasmonic routing [16], particle cleaning [17], and supercontinuum generation [11]. Recently, the observation of electron Airy beam has also been reported [18].

The dynamics of truncated Airy pulses launched into a fiber close to its zero-dispersion point under the action of second- and third-order dispersion (TOD) has been investigated [19-21]. It was demonstrated that, additional TOD of the same sign that its second order counterpart leads to variations in curvature of trajectory [19]. On the other hand, when the pulse dynamics is governed by the TOD of an opposite sign to the second-order dispersion (SOD) term, the Airy pulse reaches the tight-focusing point, then undergoes an inversion, and finally continues to propagate with an opposite acceleration [20]. At the focal point, the pulse is concentrated in a very narrow and intense light spot (hot spot). Under the action of SOD and TOD with comparable strengths, the focal point extends into a finite area, from which the pulse re-emerges with its acceleration reversed [20]. Interestingly, reversing the acceleration of Airy waves in spatial domain was recently demonstrated by applying nonlinear three wave mixing process in an asymmetrically-poled nonlinear photonic crystal [22].

It was found that the linear propagation of the truncated Airy wave, which is suitable for a practical realization, in a fiber with SOD leads to a final diversion of the wave [8]. Very recently a scheme based on dispersion management (DM) [23] was proposed for the increase of Airy pulse longevity [24]. TOD of the same sign as SOD was also taken into account.  In general, any initial field distribution can be recovered using the DM method and limitations start to rise only in nonlinear regime.

Here, we analyze and compare possibilities to regenerate truncated Airy pulses by the application of a DM scheme consisting of alternating fiber sections with opposite signs of the second order dispersion term similar to [24] with the regeneration possibilities in the new scheme with alternation of the both two leading dispersion terms. In the scheme with the alternation of both the second and the third dispersion terms it will be demonstrated that the acceleration of the Airy pulse will experience multiple reversals which are accompanied by the creation of multiple hot spots in a course of the propagation in the dispersion managed fiber link. The later scheme will reveal a long lasting propagation with multiple tight light concentration hot spots in a course of the propagation. This regime of a tight pulse compression persists in a presence of strong fiber Kerr nonlinearity and provides a possibility to describe the system performance in a quantitative way as a function of the injected light peak power.
\section{The model}

The evolution of truncated Airy pulses in lossless fiber links with alternating signs of SOD and TOD of opposite signs is considered. The evolution is governed by the normalized nonlinear Schrodinger equation:
\begin{equation}
i\phi _{\xi }+(-1)^{n}(1/2)\phi _{TT}+(-1)^{n+1}(i/6)\varepsilon \phi
_{TTT}+|\phi |^{2}\phi =0 \label{eq}
\end{equation}%
An odd integer index-$n$ represents sections with negative SOD and positive
TOD, while an even integer index-n stands for fiber sections with positive
SOD and negative TOD. The normalized distance  is given by $\xi =z|\beta
_{2}|/T_{0}^{2}$ and the relative TOD strength by the in our scheme always
negative parameter  $\varepsilon =\beta _{3}/(\beta _{2}T_{0})$, where $%
\beta _{2}$ and $\beta _{3}$ are the second- and third-order dispersion
parameters of the fiber sections, and $T_{0}$ is the pulse width. The Kerr
nonlinearity is represented by the last term.

In general the SOD parameter can be very small if one operates near the zero dispersion point of a fiber. In case of a very steep slope of the dispersion a relative TOD-SOD parameter - $\varepsilon$ can take large values. Strictly speaking a generalized Nonlinear Schroedinger equation [25] including the Raman and shock terms would describe the model more precisely in such cases. Still the Raman term maybe excluded in fibers such as hollow core photonic crystal fibers  filled with Raman inactive gas [26] and the shock term doesn't change the system behavior in a significant manner as direct numerical simulations including this term demonstrated.

\section{Dynamics in quasi-linear regime}

In case of the negligible TOD parameter (for example for pulses with long duration) the third term in Eq. (1) can be dropped and a classical DM scheme with compensation of only the SOD applies. An example for the application of dense DM scheme [27] to the Airy pulse is presented in Fig. 1(a) for 5 representative scheme periods with fibers connections of $\xi =4$. We can see that the pulse shape is periodically recovered during the propagation.
For shorter pulses usually the role of TOD increases and we have to take into account its influence on the dynamics of Airy pulses [19-21, 24]. In practice, if we make fiber connections too frequent, losses associated with these connections will play a stronger role. On the other hand if we connect compensating section too far from the preceding one, the pulse will exceed its initial temporal slot. We chose an optimal regime for the system performance to preserve the original pulse width working with each fiber section's length:
\begin{equation}
L=2/\varepsilon \label{eq}
\end{equation}%
 and thus one period of DM scheme is 2L.
Practically, an Airy pulse can be produced by adding a cubic phase to the spectrum of a simple Gaussian i.e., the input for the Eq. (1) will be $\varphi _{0}(\omega )=\exp (-\alpha \omega ^{2}+i\omega ^{3}/3)$ with $\alpha>0$. In the low power propagation regime, the nonlinear term in Eq. (1) can be dropped and the input in frequency representation can be substituted into Eq. (3), that is transformed Eq. (1).
\begin{equation}
\psi _{\xi }=(i/2)sgn(\beta _{2})\omega ^{2}\psi +(1/6)\varepsilon
\omega ^{3}\psi \label{eq}
\end{equation}%
After the inverse Fourier transform the analytic solution for the pulse evolution in odd and even fiber links reads:
\begin{gather}
|\varphi (\xi \prime ,T)|^{2}=2\pi \theta ^{2/3}|Ai[\theta
^{1/3}(T-0.25\theta \xi \prime ^{2}+  \notag \\
+\alpha ^{2}\theta +i\alpha \theta \xi
\prime )]|^{2}\*exp (2\alpha \theta T+\frac{4}{3}\alpha ^{3}\theta
^{2}-\theta ^{2}\alpha \xi \prime ^{2}),  \label{4(a)}
\end{gather}%
for odd fiber sections with $\xi \prime =\xi -2nL,$
\begin{gather}
|\varphi (\xi \prime \prime ,T)|^{2}=2\pi \theta ^{2/3}|Ai[-\theta
^{1/3}(T-0.25\theta \xi \prime \prime ^{2}+  \notag \\
+\alpha ^{2}\theta -i\alpha \theta
\xi \prime \prime )]|^{2}\exp (-2\alpha \theta T-\frac{4}{3}\alpha
^{3}\theta ^{2}+\theta ^{2}\alpha \xi \prime \prime ^{2}),  \label{4(a)}
\end{gather}%
for compensating even fiber sections with $\xi\prime\prime =\xi-(2n+1)L$.
The parameter $\theta$ is given by $\theta =1/(0.5\varepsilon \xi +1)$. In the center of each of the fiber sections with $0.5\varepsilon \xi =1$, the pulse reaches its most compressed state and the pulse profile is Gaussian and given by [20]:
\begin{equation}
|\varphi (T)|^{2}=\frac{1}{2\sqrt{\alpha ^{2}+1/\varepsilon ^{2}}}\exp (-%
\frac{T^{2}}{2\alpha +2/(\alpha \varepsilon ^{2})}) \label{eq}
\end{equation}%

The given analytic solutions were compared to the results of direct numerical simulations of Eq. (1) and a complete agreement was found.
The pulses are launched as [8]:
 \begin{equation}
\varphi _{0}=AAi(T)\exp (\alpha T)  \label{eq}
\end{equation}%
 with the truncation parameter $\alpha>0$. Fig. 1(b) demonstrates the evolution of low power Airy pulses with amplitude $A=0.1$ for the case of joint action of SOD and TOD of equal strengths. In this regime, near the middle of each section of the link the Airy pulse reaches a finite area of recombination and afterwards experiences an acceleration reversal. The size of the area depends on the relative strength of the SOD and TOD coefficients.

Next, we consider the dynamics of the low peak power Airy pulses under the dominant action of TOD over SOD which is relevant when pulses are launched very close to the zero dispersion point of the fiber (Fig. 1(c)). In the regime of TOD domination the recombination area converges into a very short temporal slot, featuring hot spots with tight focusing in the time domain with all the pulse's energy concentrated in it.
\begin{figure}[tbp]
\centering
\includegraphics[width=1\columnwidth]{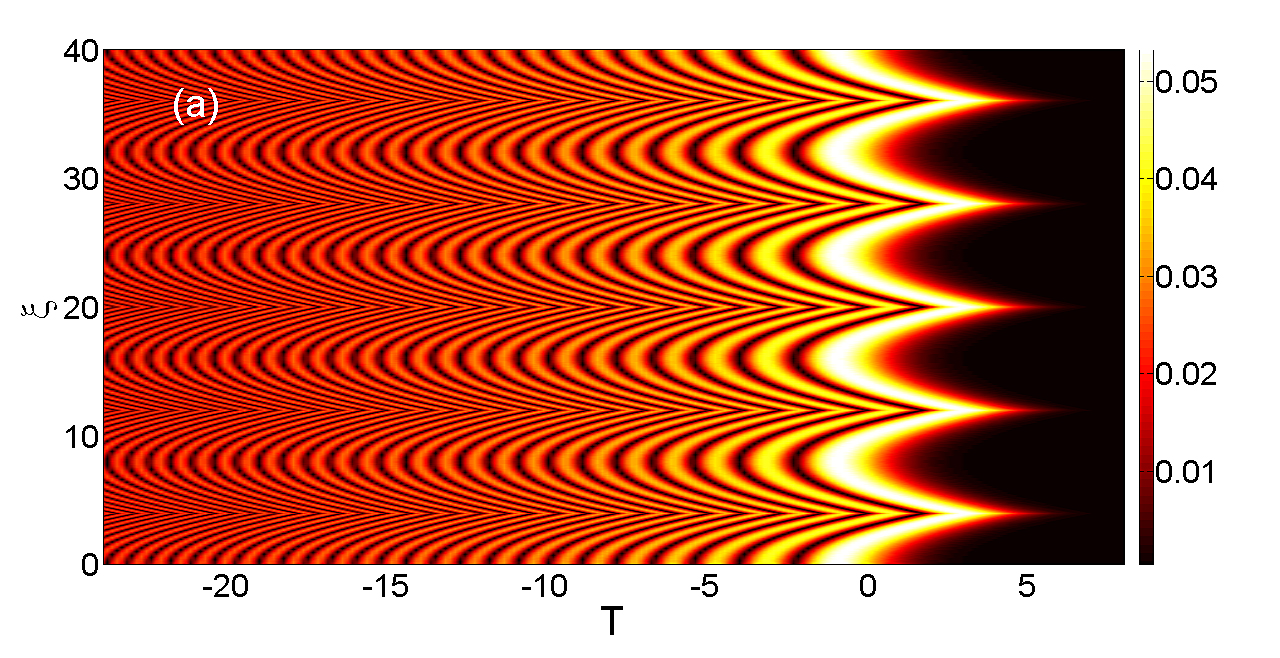}
\includegraphics[width=1\columnwidth]{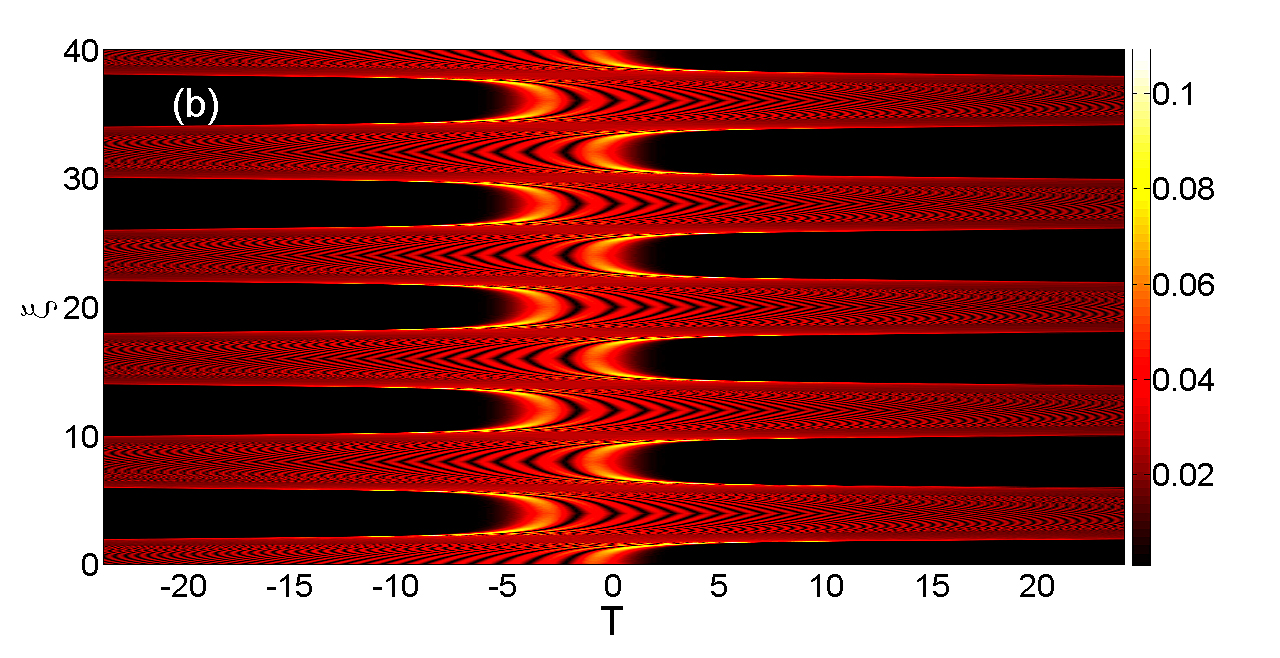}
\includegraphics[width=1\columnwidth]{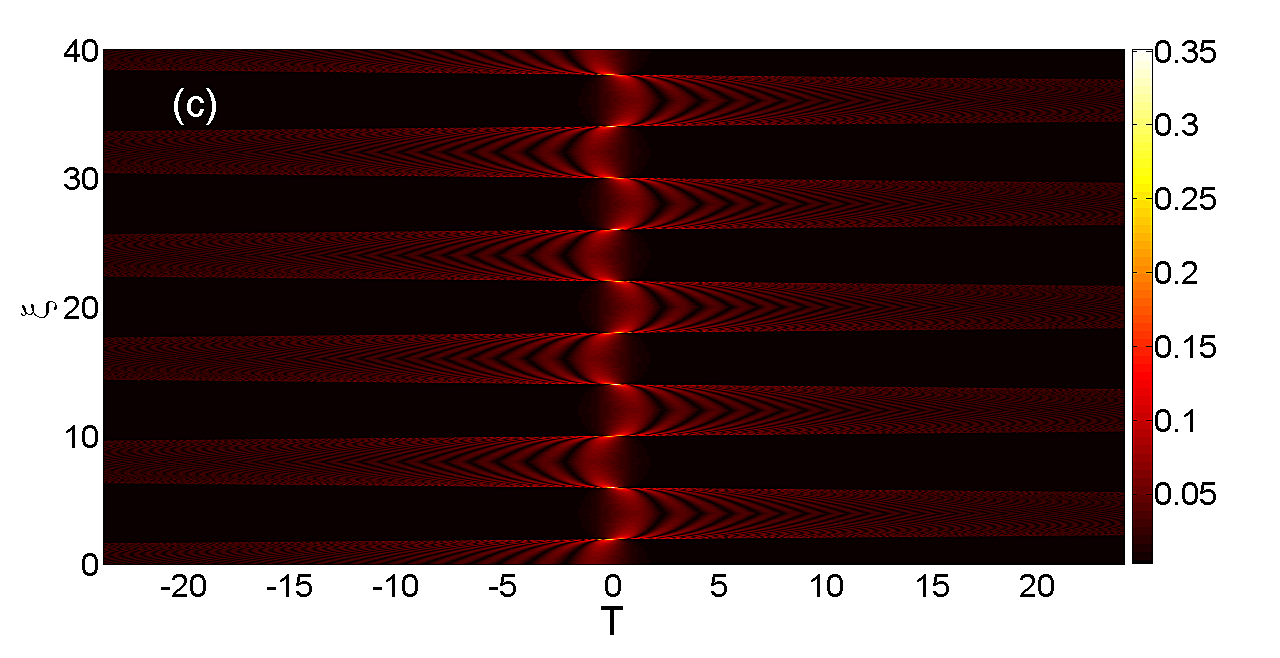}
\caption{(Color online)(a) The dynamics of Airy pulse in the quasi-linear regime in a DM with alternation of SOD only ($\epsilon = 0$). (b)  The dynamics of Airy pulses in 5 periods of the DM scheme under the action of equal strength SOD and TOD, with $\epsilon = -1$ in Eq. (1). (c) The dynamics under the action of strongly dominant of TOD with $\epsilon = -1000$.  Here and below, all results are presented in normalized units. Here and in the next figure the absolute value of field is shown instead of the intensity for better visibility of the low power regions of the pulse.}
\end{figure}
As we can see the DM scheme provides a successful restoration of Airy pulses in low power regime that in practice is only limited by the fiber's attenuation.
\section{Dynamics in the strongly nonlinear regime.}
Launching pulses in the high-power strongly nonlinear regime leads to some deterioration of the system performance while the main effects remain qualitatively valid. Fig.2 (a,b) demonstrates distortion with propagated distance of the Airy pulse launched in fiber link as in Fig. 1, but with an amplitude A=3 (30 times higher than in case of Fig. 1(a,b)). Fig.2 (a) demonstrates the distorted dynamics in nonlinear regime using the regular DM scheme with the compensation of SOD only, while the Fig.2 (b) demonstrates the same pulse evolution in a scheme with management of both SOD and TOD.
 \begin{figure}[tbp]
\centering
\centering\includegraphics[width=1.0\columnwidth]{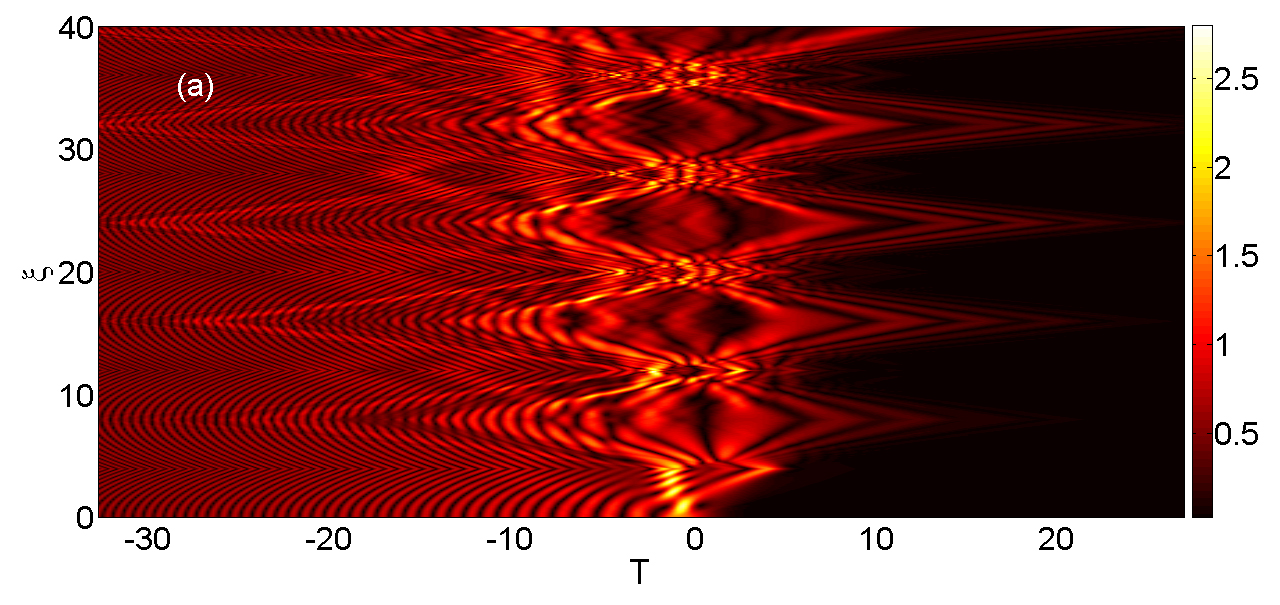}
\centering\includegraphics[width=1.0\columnwidth]{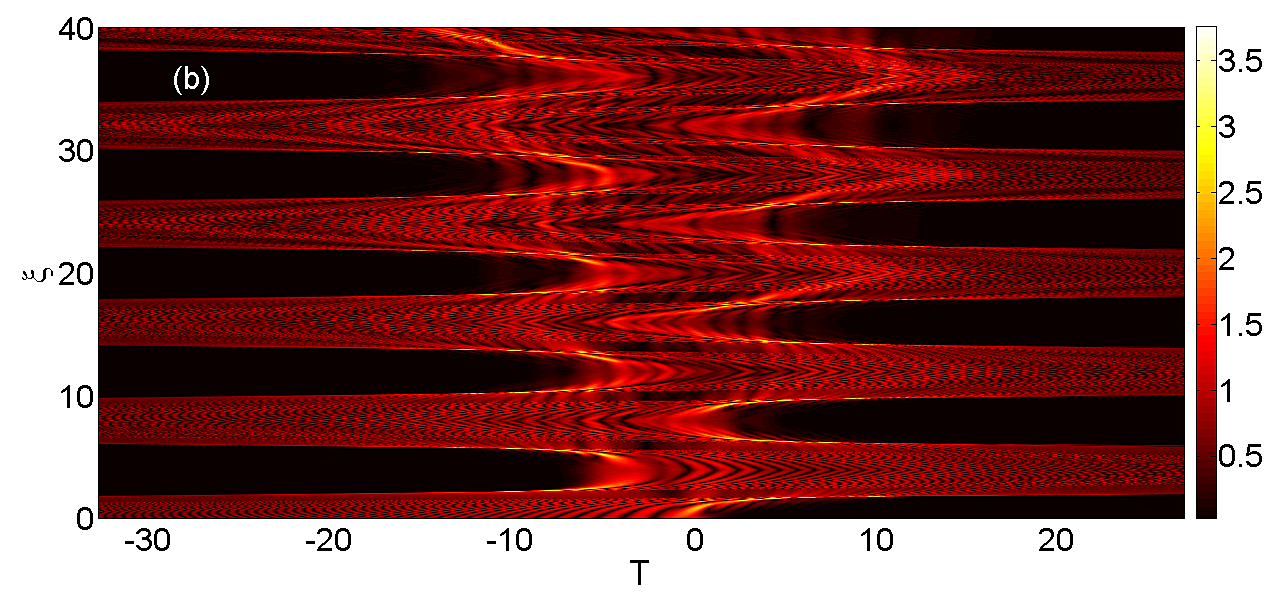}
\centering\includegraphics[width=1.0\columnwidth]{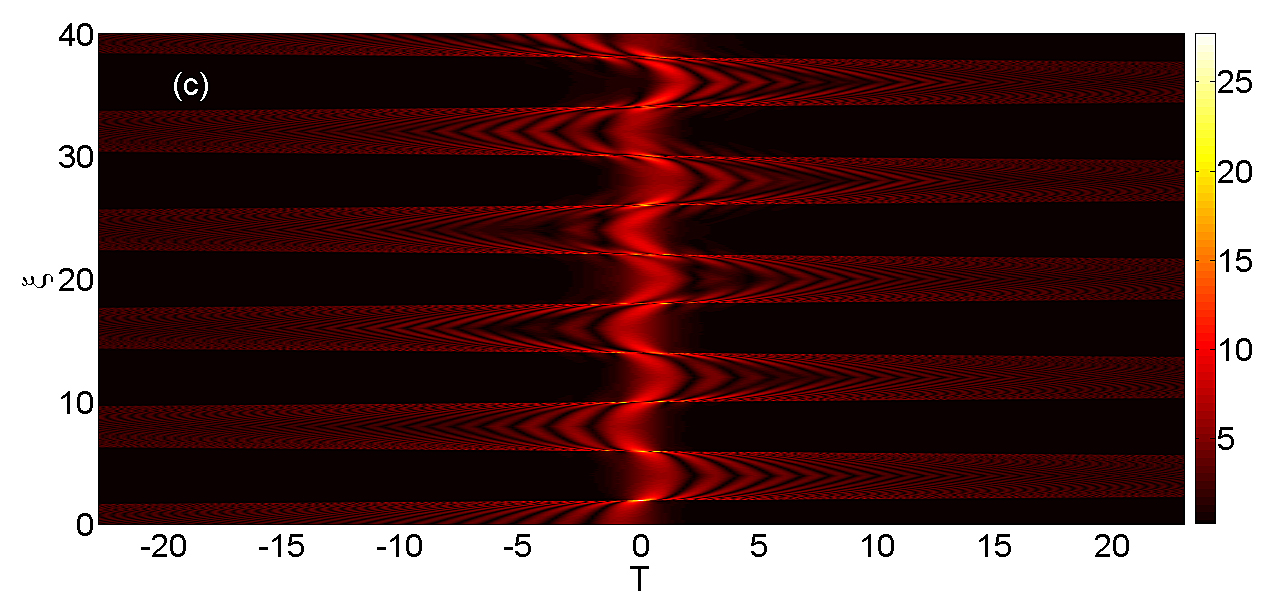}
\caption{(Color online) The dynamics of Airy pulses in the strongly nonlinear regime  with the amplitude A=3. (a)The dynamics for 5 periods of the DM scheme based only on SOD compensation (b) for 5 periods of the DM scheme under the equal action of SOD and TOD with $\epsilon = -1$. (c) for 20 periods of the DM scheme under the dominance of TOD with $\epsilon = -1000$ and the amplitude A=10.}
\end{figure}

One can reduce the negative effect of nonlinearity, choosing the central wavelength of the pulse to be near the zero dispersion point of the fiber, which means working with a high parameter - $\epsilon$. Fig. 2(c) demonstrates such a dynamics in highly nonlinear regime of Airy pulse in fiber link with $\epsilon = -1000$ as in Fig. 1(c) with an amplitude A=10, that is 100 times higher than that pertaining to Fig. 1(c).
 It is shown, that the pulse successfully propagates 5 DM periods only with some partial deformation.

 For such a fiber link with pronounced dominance of TOD one can estimate the dependence of system performance deterioration on nonlinearity effect by studying the pulse's intensity enhancement deterioration with the distance for inputs with various amplitudes. This can be best observed at the focal points, where the pulse becomes mostly compressed. Fig. 3 demonstrates evolution of the peak intensity of Airy pulse with amplitude A=1 (a) and A=2 (b) and with the truncation $\alpha = 0.001$ in fiber link with 40 DM periods with $\epsilon = -100$. One can clearly see from Fig. 4, that 80 light hot spots are created within the propagation of the 40 DM periods. For the case of A=1 presented in Fig. 3(a) the peak power ratio between its value at the focal points and at the input is partially reduced from 37.25 to 17.67 in a course of propagation of the 40 DM periods. In case of stronger influence of nonlinearity further degradation of pulse compression and power enhancement at focal po
 ints is expected.
 \begin{figure}[tbp]
\centering
\includegraphics[width=0.9\columnwidth]{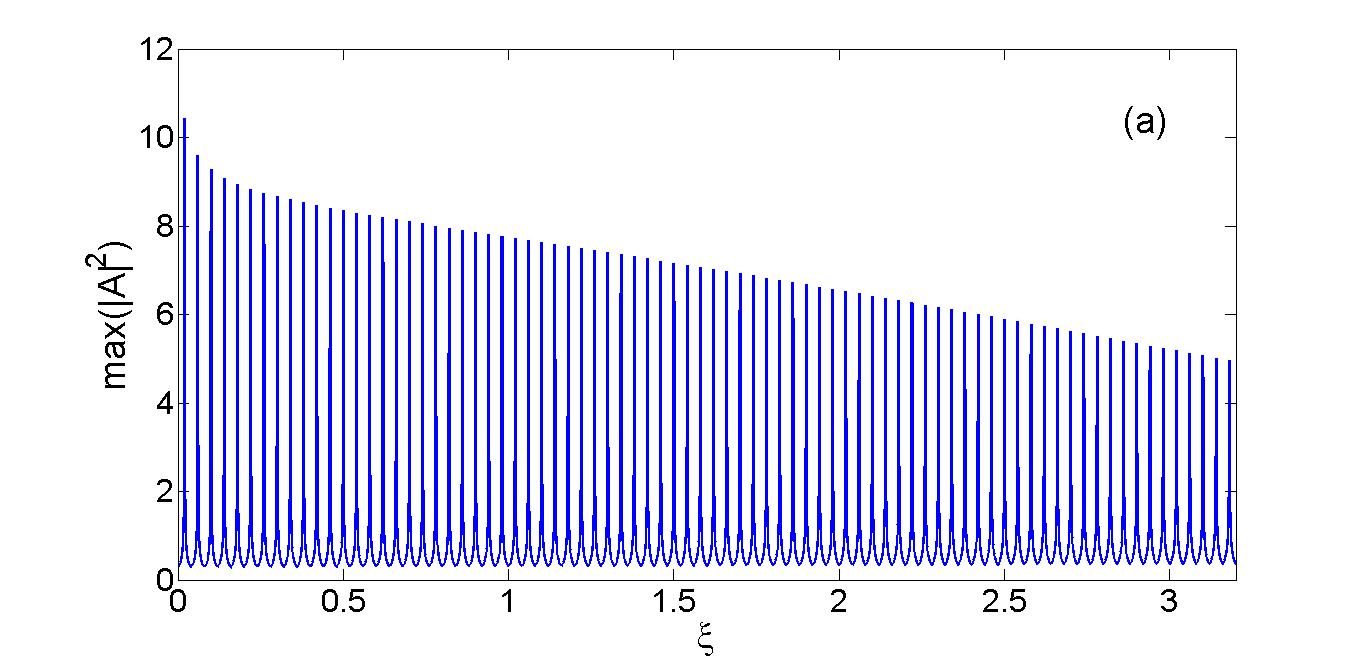}
\includegraphics[width=0.9\columnwidth]{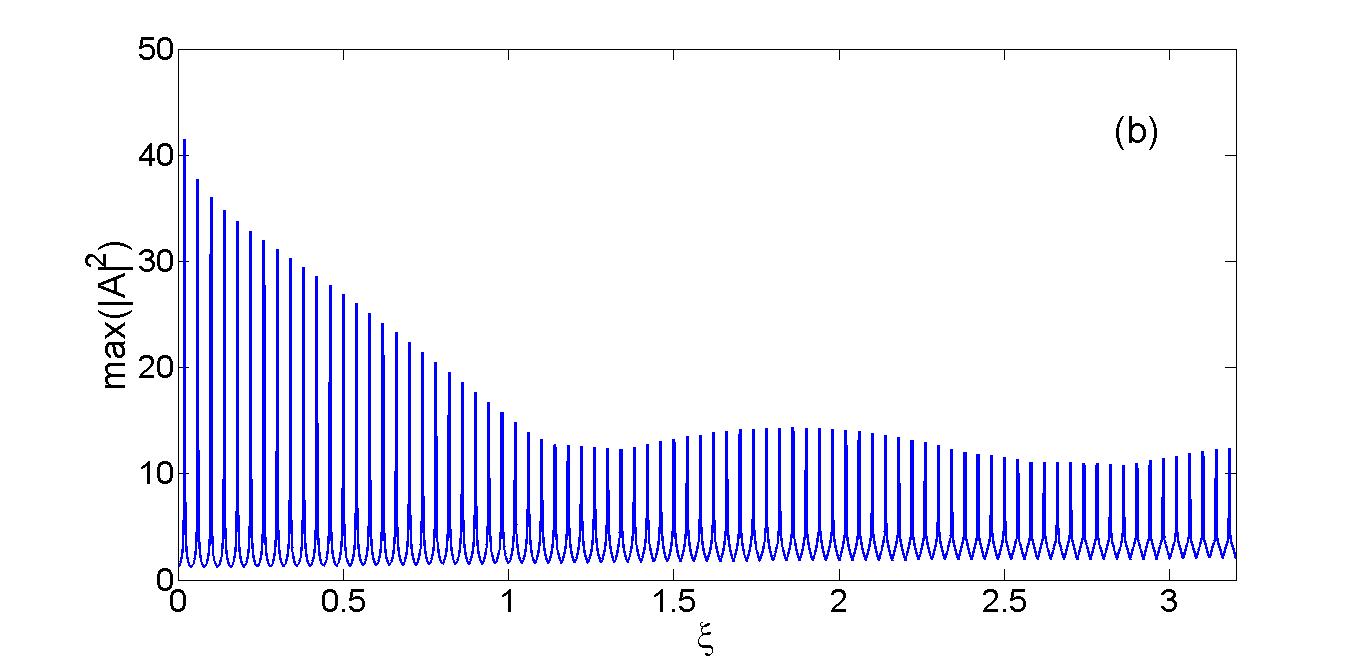}
\caption{(Color online) Evolution of the peak power of the Airy pulse in the strongly nonlinear regime for 40 periods of the DM scheme under the dominant action of TOD with $\epsilon = -100$ and an amplitude of A=1 in (a) and of A=2 in (b).}
\end{figure}
Fig. 3(b) reveals the much faster decrease in peak power enhancement for the input amplitude A=2 within first 15 DM periods up to the propagated distance about $\xi=1.2$. Further the decrease becomes less dramatic featuring oscillating behavior. Launching even more intensive pulses may also lead to solitons shedding out of the Airy pulse structure [20, 28]. Fig. 4(a) demonstrates the deterioration of the peak power enhancement with each of the 20 propagated DM periods with $\epsilon = -100$. The two cases of dynamics with the injected amplitudes A=1, 2, corresponding to Fig. 3, are represented by the blue and red curves respectively. Also additional cases of lower amplitude A=0.5 (the green curve) and higher amplitude A=5 (the black curve) are demonstrated.

One can achieve a significant improvement in the system performance working even closer to the zero dispersion point of the fiber with the larger value of e. Fig. 4(b) demonstrates such an improvement for the same pulses dynamics as in Fig. 4(a), but with the increased value of the relative TOD strength parameter- $\epsilon = -1000$.

 \begin{figure}[tbp]
 \centering
\includegraphics[width=0.8\columnwidth]{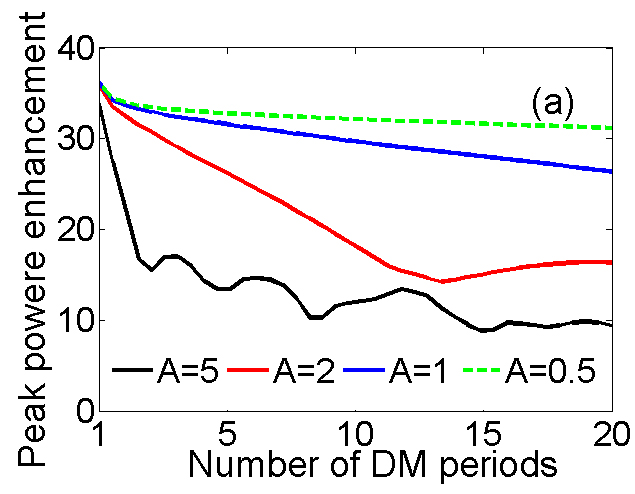}
\includegraphics[width=0.81\columnwidth]{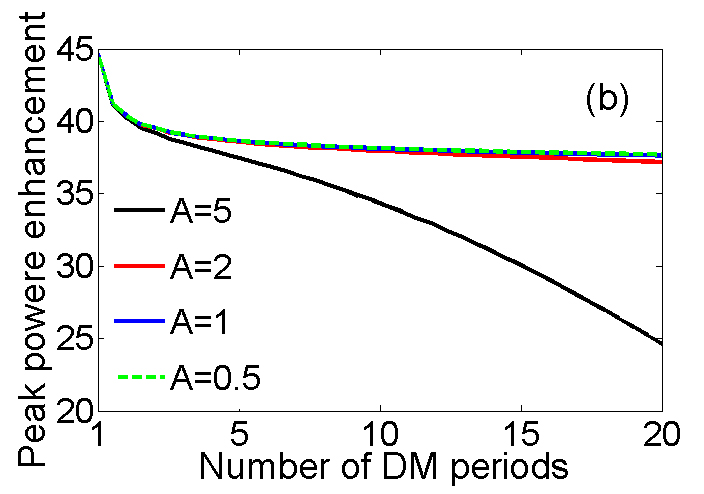}
\caption{(Color online) Reduction of the peak power enhancement in nonlinear propagation regimes with various pulse's peak intensities for 20 periods of the DM scheme under the dominant action of TOD with $\epsilon = -100$ (a) and with $\epsilon = -1000$ (b). The green, blue, red and black curves stand for the cases with A=0.5, 1, 2, and 5 respectively.}
\end{figure}

 The cases with the injected amplitudes A=0.5, 1, 2 almost overlap demonstrating small degradation. The case with a very large amplitude A=5 demonstrates significant decrease in the pulse's peak power enhancement of about 45 percents, after the propagation of 20 DM periods, which is still much better than its counterpart in Fig. 4(a) representing 74$\%$ performance deterioration.

\section{Conclusion}

We have demonstrated a method for the regeneration of truncated self-accelerating Airy pulses in fiber links using the famous dispersion management technique in which fiber sections with opposite signs of second and third order dispersion parameters are periodically alternating. In the low-power linear regime robust propagation of Airy pulses is limited only by fiber losses and the dynamics of these pulses is fully described analytically.
Comprehensive numerical studies demonstrate the qualitative validity of the main regeneration effect with good pulse shape maintenance also in the nonlinear regime. The system performance in the presence of strong nonlinear effect can be further optimized by launching the input closer to the zero dispersion point of fibers which corresponds to increasing the relative strength of the third order dispersion.

\section{Acknowledgments}
The Authors gratefully acknowledge support provided by the Deutsche Forschungsgemeinschaft
(DFG) via the Research Training Group (GRK 1464), and computing time provided by
PC2 (Paderborn Center for Parallel Computing).

\end{document}